\documentclass{emulateapj}
\submitted{{\it Submitted for publication in ApJL}}
\usepackage{multirow,color,wrapfig}
\usepackage {graphicx}
\usepackage{graphics}
\usepackage[dvips]{epsfig}

\newcommand{\beq}{\begin{eqnarray}}  
\newcommand{\eeq}{\end{eqnarray}}

\newcommand{\ly}{{\ifmmode{{\rm Ly}\alpha}\else{Ly$\alpha$}\fi}}
\newcommand{\hMpc}{{\ifmmode{h^{-1}{\rm Mpc}}\else{$h^{-1}$Mpc}\fi}}  
\newcommand{\hGpc}{{\ifmmode{h^{-1}{\rm Gpc}}\else{$h^{-1}$Gpc}\fi}}  
\newcommand{\hmpc}{{\ifmmode{h^{-1}{\rm Mpc}}\else{$h^{-1}$Mpc}\fi}}  
\newcommand{\hkpc}{{\ifmmode{h^{-1}{\rm kpc}}\else{$h^{-1}$kpc}\fi}}  
\newcommand{\hMsun}{{\ifmmode{h^{-1}{\rm {M_{\odot}}}}\else{$h^{-1}{\rm{M_{\odot}}}$}\fi}}  
\newcommand{\hmsun}{{\ifmmode{h^{-1}{\rm {M_{\odot}}}}\else{$h^{-1}{\rm{M_{\odot}}}$}\fi}}  
\newcommand{\Msun}{{\ifmmode{{\rm {M_{\odot}}}}\else{${\rm{M_{\odot}}}$}\fi}}  
\newcommand{\msun}{{\ifmmode{{\rm {M_{\odot}}}}\else{${\rm{M_{\odot}}}$}\fi}}

\newcommand{\rand}{{\ifmmode{{\mathcal{R}}}\else{${\mathcal{R}}$ }\fi}}  

\newcommand{\reff}{{\ifmmode{r_{\mbox{\tiny eff}}}\else{$r_{\mbox{\tiny eff}}$}\fi}}

\begin{document}

\title{Quantifying and controlling biases in dark matter halo concentration estimates}
\author{
  C. N. Poveda-Ruiz \thanks{cn.poveda542@uniandes.edu.co}$^{1}$
  J. E. Forero-Romero \thanks{je.forero@uniandes.edu.co}$^{1}$
  J. C. Mu\~noz-Cuartas \thanks{juan.munozc@udea.edu.co}$^{2}$
}

\affil{
$^1$Departamento de F\'{i}sica, Universidad de los Andes, Cra. 1
No. 18A-10, Edificio Ip, Bogot\'a, Colombia\\
$^2$Instituto de F\'{\i}sica - FCEN, Universidad de Antioquia, Calle
67 No. 53-108, Medell\'{\i}n, Colombia
}

\begin{abstract}
We use bootstrapping to estimate the bias of concentration estimates
on N-body dark matter halos as a function of particle number.  We find
that algorithms based on the maximum radial velocity and radial
particle binning tend to overestimate the concentration by $15\%-20\%$
for halos sampled with $200$ particles and by $7\%$-$10\%$ for halos
sampled with $500$ particles.  To control this bias at low particle
numbers we propose a new algorithm that estimates halo concentrations
based on the integrated mass profile.  The method uses the full
particle information without any binning, making it reliable in cases
when low numerical resolution becomes a limitation for other methods.
This method reduces the bias to $< 3\%$ for halos sampled with
$200$-$500$ particles.  The velocity and density methods have to use
halos with at least $\sim 4000$ particles in order to keep the biases
down to the same low level.  We also show that the mass-concentration
relationship could be shallower than expected once the biases of the
different concentration measurements are taken into account.  These
results show that bootstrapping and the concentration estimates based
on the integrated mass profile are valuable tools to probe the
internal structure of dark matter halos in numerical simulations.
\end{abstract}


\section{Introduction}
\label{sec:introduction}
In the current structure formation paradigm the properties of galaxies
are coupled to the evolution of their dark matter (DM) hosting halo.
In this paradigm the sizes and dynamics of galaxies are driven by
the halo internal DM distribution. 

The internal DM distribution in a halo is usually parameterized
through the density profile.  
In a first approximation this profile is spherically symmetric; the
density only depends on the radial coordinate.  
One of the most popular radial parameterizations is the
Navarro-Frenk-White (NFW) profile \citep{NFW}.   
This profile can be considered as universal \citep{Navarro2010}, assuming that 
one is not interested in the very central region where galaxy
formation takes place, and where the effects of baryon physics on the
DM distribution are still unknown. 
This profile is a double power law in radius, where the transition
break happens at the so-called scale radius, $r_s$.  
The ratio between the scale radius and the halo virial radius $R_v$ is
known as the concentration $c=R_v/r_s$.

The concentration of the NFW profile provides a conceptual framework
to study simulated DM halos as a function of redshift and
cosmological parameters.  
Numerical studies
\citep{Neto2007,Maccio2008,Duffy2008,Munoz2011,Prada2012,Ludlow2014,Ludlow2016,Klypin2016}
summarized their results through the mass-concentration
relationship; that is, the distribution of concentration values at a
fixed halo mass and redshift.  
The success of such numerical experiments rests on a reliable
algorithm to estimate the concentration.  
Such an algorithm should provide unbiased results and must be robust
when applied at varying numerical resolution. 

There are two established algorithms to estimate the concentration
parameter.
The first method takes the halo particles and bins them into
logarithmic radii to estimate the density in each bin, then it
proceeds to fit the density as a function of the radius.  
A second method uses an analytic property of the NFW profile that
relates the maximum of the ratio of the circular velocity to the
virial velocity, $V_{\rm circ}$/$V_{\rm   vir}$.  The concentration
can be then found as the root of an algebraic equation dependent on
this maximum value. 

The first method is straightforward to apply but presents two
disadvantages.  First, it requires a large number of particles in
order to have a proper density estimate in each bin.  This makes the
method robust only for halos with at least $10^2$ particles.  The
second problem is that there is not a way to estimate the optimal
radial bin size, different choices may produce different results for
the concentration.

The second method solves the two problems mentioned above.  
It works with low particle numbers and does not involve data binning.  
However, it effectively takes into account only a single data point
and discards the rest of the data.
Small fluctuations on the maximum can yield large perturbations on the
estimated concentration parameter.

In this letter we use bootstrapping to estimate the bias and standard
deviation on the concentration estimates as a function of particle
number. 
We show that the two standard methods to estimate concentrations have
increasing biases for decreasing particle numbers.  

This motivates us to present a third alternative based on fitting the
integrated mass profile.  
This approach has two advantages with respect to the above mentioned methods.  
It does not involve any data binning and does not throw away data points.  
This translates into a robust estimate even at low resolution/particle
numbers.   
Furthermore, since the method does not require any binning, there is no need to
tune numerical parameters.  
This is a new independent method to estimate the
concentration parameter.

\section{Basic properties of the NFW density profile}
\label{sec:basics}

Let us review first the basic properties of the NFW density profile.
This shall help us to define our notation.

\subsection{Density profile}

The NFW density profile can be written as

\begin{equation}
\rho(r) = \frac{\rho_c\delta_c}{r/r_s(1+r/r_s)^2},
\label{eq:definition}
\end{equation}
where $\rho_c\equiv 3H^2/8\pi G$ is the Universe critical density, $H$
is the Hubble constant, $G$ the universal gravitational constant,
$\delta_c$ is the halo dimensionless characteristic density and $r_s$
is the scale radius.  This radius marks the point where the
logarithmic slope of the density profile is equal to -2, the
transition between the power law scaling $\rho\propto r^{-1}$ for
$r<r_s$ and $\rho\propto r^{-3}$ for $r>r_s$.

We define the virial radius of a halo, $r_v$, as the boundary of the
spherical volume that encloses a density of $\Delta_h$ times the mean
density of the Universe.  The corresponding mass $M_{v}$, the virial
mass, can be written as $M_{v} = \frac{4\pi}{3}\bar{\rho}\Delta_h
r_v^3$.  From these virial quantities we define new dimensionless
variables for the radius and mass $x\equiv r/r_v$ and $m\equiv
M(<r)/M_v$.

In this letter we use $\Delta_h=740$, a number roughly corresponding to
200 times the critical density at redshift z=0.

\subsection{Integrated mass profile}

From these definitions we can compute the total mass enclosed inside a
radius $r$:
\begin{equation}
M(<r) = 4\pi\rho_c\delta_c  r_s^3\left[\ln \left
  (\frac{r_s+r}{r_s}\right) - \frac{r}{r_s+r}\right],
\end{equation}
or in terms of the dimensionless mass and radius variables

\begin{equation}
m(<x) =
\frac{1}{A}\left[\ln\left(1+xc\right)-\left(\frac{xc}{xc+1}\right)\right],
\label{eq:profile}
\end{equation}
where
\begin{equation}
A=\ln\left(1+c\right)-\left(\frac{c}{c+1}\right),
\end{equation}
and the parameter $c$ corresponds to the concentration $c\equiv
r_v/r_s$.

From this normalization and for later convenience we define the
following function
\begin{equation}
f(x) = \ln\left(1+x\right)-\left(\frac{x}{x+1}\right).
\label{eq:f_NFW}
\end{equation}
The most interesting feature of Eq. (\ref{eq:profile}) is that the
concentration is the only free parameter to describe the integrated
mass profile.
 
\subsection{Circular velocity profile}

It is also customary to express the mass of the halo in terms of the
circular velocity $V_{c}=\sqrt{GM(<r)/r}$.  From this we can define a
new dimensionless circular velocity $v(<x)\equiv
V_{c}(<r)/V_{c}(<r_v)$, using the result in Eq. \ref{eq:profile} we
have:

\begin{equation}
v(<x)=\sqrt{\frac{1}{A}\left[\frac{\ln\left(1+xc\right)}{x}-\frac{c}{xc+1}\right]},
\end{equation}
This normalized profile always shows a maximum provided that the
concentration is larger than $c>2$.  It is possible to show that for
the NFW profile the maximum is provided by

\begin{equation}
\mathrm{max}(v(<x)) = \sqrt{\frac{c}{x_{\rm max}}\frac{f(x_{\rm
      max})}{f(c)}},
\label{eq:max_v}
\end{equation}
where $x_{\rm max}=2.163$ \citep{Klypin2016} and the function $f(x)$
corresponds to the definition in Eq. (\ref{eq:f_NFW}).

\begin{figure*}
\begin{center}
  \includegraphics[width=0.49\textwidth]{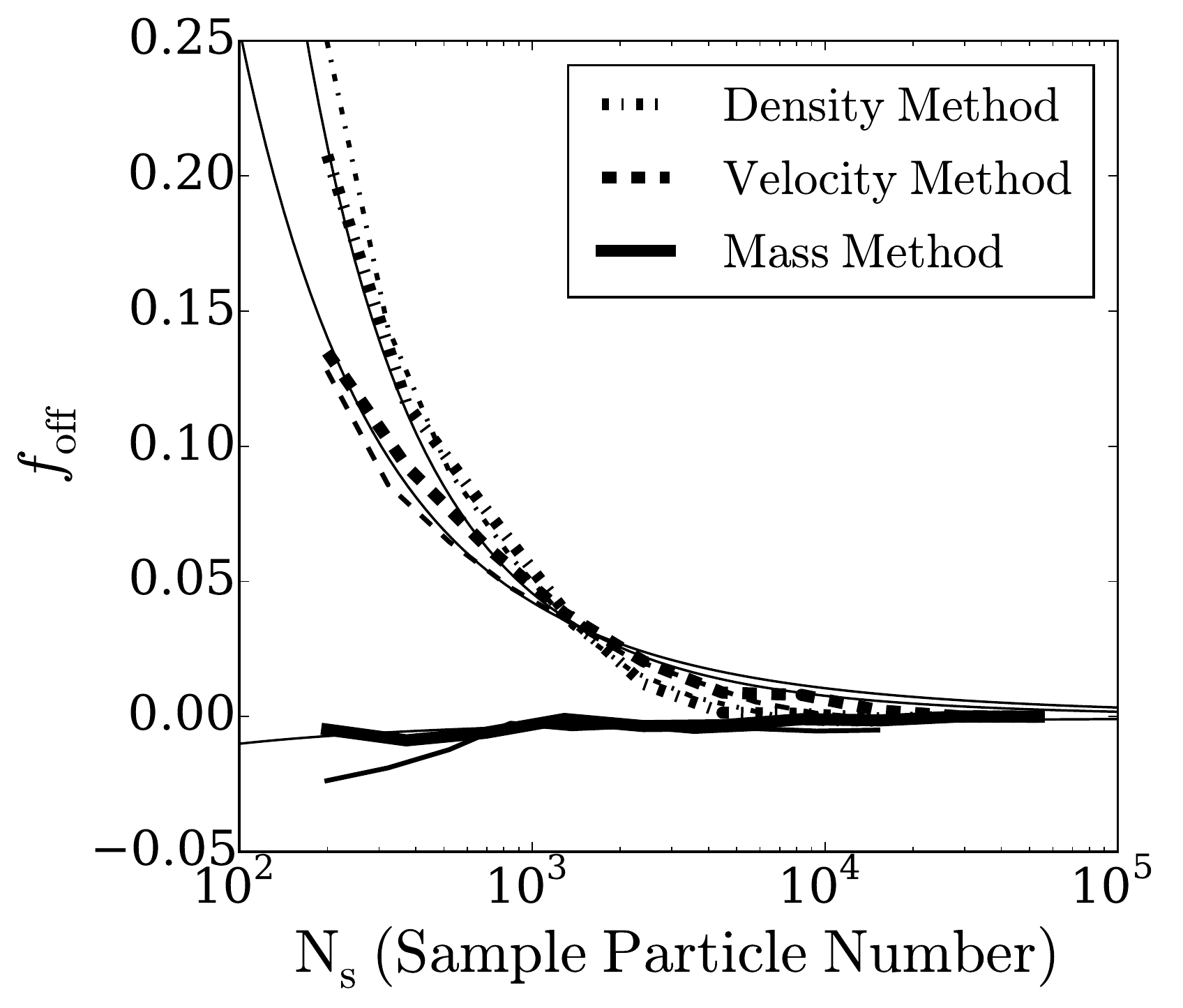}
  \includegraphics[width=0.48\textwidth]{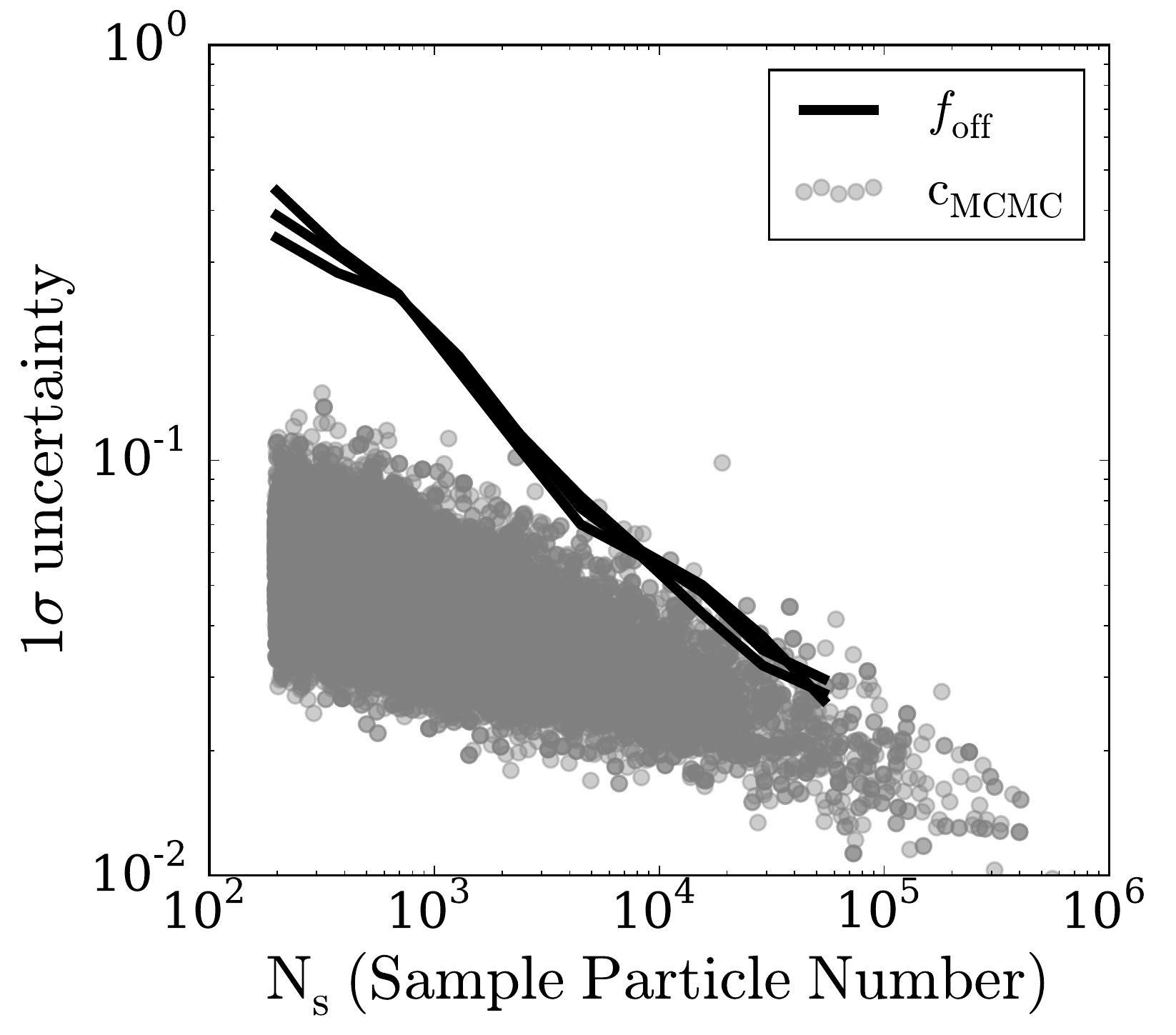}
\end{center}
\vspace{-0.5cm}
\caption{\emph{Left panel}. Bias estimated via bootstrapping on the
  concentration as a function of particle number.
  Thick (thin) lines correspond to massive halos in the Bolshoi (Via
  Lactea) simulation.
  The density method noticeably overestimates the concentration up to
  a factor of $20\%$, 
  while the new method only underestimates the concentrations by less
  than $3\%$.
  \emph{Right panel.}
  1$\sigma$ uncertainties on the bootrsapped halos (lines) and the MCMC
  uncertainties on the concentration estimates for each halo using the
  integrated mass method (circles).   
  Lines show the width between the 14 and 86 percentiles
  of the $f_{\rm off}$ distribution at fixed particle number. 
  The lines include the results for the three methods using Bolshoi data.
  To allow a fair comparison agains $f_{\rm off}$, the MCMC uncertainty has
  been normalized by the preferred concentration value for each halo. 
  \label{fig:downsampling}}
\end{figure*}

\section{Methods to estimate the concentration from N-body simulations}
\label{sec:method}

\subsection{Estimates from the density and velocity profiles}

To date, there are two standard methods to estimate concentrations in
DM halos extracted from N-body simulations.  The first method
takes all the particles in the halo and bins them in the logarithm of
the radial coordinate from the halo center.  Then, it estimates the
density in each logarithmic bin.  At this point is possible to make a
direct fit to the density as a function of the radial coordinate.
This method has been broadly used for more than two decades to study
the mass-concentration-redshift relation of DM halos. 
A second method uses the circular velocity profile.  It finds the
value of $x$ for which the normalized circular velocity $v(<x)$ shows
a maximum.  Using this value it solves numerically for the
corresponding value of the concentration using Eq. (\ref{eq:max_v}).

\subsection{Estimate from the integrated mass profile}

Here we propose a new method to estimate the concentration. 
It uses the integrated mass profile defined in
Eq. (\ref{eq:profile}).  We build it from N-body data as follows.
First, we define the center of the halo to be at the position of the
particle with the lowest gravitational potential. Then we rank the
particles by their increasing radial distance from the center.  From
this ranked list of $i=1,N$ particles, the total mass at a radius
$r_i$ is $M_i=i\times m_p$, where $r_i$ is the position of the $i$-th
particle and $m_p$ is the mass of a single computational particle. 
We then divide the enclosed mass $M_i$ and the radii $r_i$ by their
virial values to finally obtain the dimensionless variables $m_i$ and
$x_i$.

Using bootstrapping data (\S \ref{sec:bootstrapping}) we find that at
a given normalized radius, $x$, the logarithm of the normalized
integrated mass, $m$, approximately follows a Gaussian distribution
with variance

\begin{equation}
\sigma_x^2 = \frac{1-x}{x}\frac{1}{N}.
\label{eq:sigma}
\end{equation}

If the integrated mass values at different radii were independent from
each other we could write a likelihood distribution as ${\cal
  L}(c|x_i)\propto \exp(-\chi^2(c,x_i)/2)$ with   

\begin{equation}
\chi^2(c,x_i)= \sum_{i=2}^{N-1}\frac{[\log m_i - \log m(< x_i;c)]^2}{\sigma_i^2},
\end{equation}
where $\sigma_i^2=\sigma_x^2(x_i)$, $m(<x_i;c)$ corresponds to the
values in Eq.(\ref{eq:profile}) at $x=x_i$ for a given value of the
concentration parameter $c$, and the $i$ index sums over all the
particles in the numerical profile. 
In this computation the particles $i=1$ and $i=N$ are discarded to
avoid divergent terms in the sum.

However, tests on the bootstrapping data show that using
$\sigma_i^2=\sigma_x^2(x_i)$, instead of the full inverse covariance
matrix, grossly overestimates $\chi^2(c,x_i)$,
providing small uncertainties around the best concentration value. 
To avoid the expensive computation and inversion of a full covariance
matrix we use the bootstrapping data to calibrate an effective
$\sigma_i^2\approx \sigma_{\rm eff}^2(x_i)$.

We impose two conditions on the approximate $\sigma_{\rm eff}^{2}$. It
must keep the dependence on $x$ that we have discovered for the
diagonal elements and must give similar curves of $\chi^{2}(c,x_i)$
vs. $c$ around the minimum as the full covariance matrix.
We found that the effective $\sigma_{\rm eff}^2$ can be approximated
as 

\begin{equation}
\sigma_{\rm eff}^2 = \frac{1-x}{x}\frac{N^{1.15}}{4.5\times 10^3}.
\end{equation}

We then use an Affine Invariant Markov Chain Monte Carlo (MCMC)
implemented in the python module {\texttt{emcee}} \citep{emcee} to sample the
likelihood function distribution.
From the $\chi^2$ distribution we find the optimal concentration value
and its associated uncertainty.   
We stress that different choices for $\sigma^2_{\rm eff}$ do not
affect the optimal concentration value, only its uncertainty.

Run-time is roughly proportional to $N$. Using a single 2.3Ghz CPU core with two
walkers over 500 steps takes $\sim 0.5$ milliseconds per halo per
particle in the halo, i.e. a halo with $N=2\times 10^{3}$ can be fit
in one second.

\section{Numerical Simulations and Halo Samples}

We use two different simulations to test our methods.
The first is the Bolshoi run, a cosmological simulation that follows
the non-linear evolution of a DM density field sampled with
$2048^3$ particles over a cubic box of $250\ \hMpc$ on a side.   The
cosmological parameters use a Hubble parameter $h=0.73$, a matter
density $\Omega_m=0.3071$ and a normalization of the power spectrum
$\sigma_8=0.82$. 
The data is publicly available at \url{http://www.cosmosim.org/}.  
Details about the structure of the database and the simulation can be
found in \citep{2011ApJ...740..102K,2013AN....334..691R}.

We use the halos located in a cubic sub-volume of $100$ \hMpc\ on a
side containing a total of $64531$ objects.
From this sample we select all the halos at $z=0$ detected with a
Friends-of-Friends (FoF) algorithm with more than 300 particles,
meaning that the masses are in the interval $4\times 10^{10}\leq
M_{\rm FoF}/\hMsun \leq 10^{14}$.  The FoF algorithm used a linking
length of $0.17$ times the mean inter-particle distance.  
This choice translates into an overdensity $\Delta_h\sim 400-700$
dependent on the halo concentration \citep{More2011}.

From this set of particles we follow the procedure spelled out in
Section \ref{sec:method} with $\Delta_h=740$ to select an spherical
region that we redefine to be our halo.  
This choice makes that the overdensities are fully included inside the
original FoF particle group.   
On the interest of providing a fair comparison against the density
method we only report results from overdensities with at least $200$
particles ($2.6\times 10^{10}$\hMsun). 

We also use public data from the Via Lactea simulation project
\citep{2008Natur.454..735D}.  
This simulation contains a single isolated halo with a virial mass
of the order of $10^{12}$\hMsun simulated using the tree code PKDGRAV
\citep{2001PhDT........21S}.  
The simulation had $\sim 2\times 10^{8}$ particles to resolve this
region.
The cosmological parameters are different from those in the Bolshoi
simulation, with a Hubble parameter $h=0.73$, a matter density
$\Omega_m=0.238$ and a normalization of the power spectrum
$\sigma_8=0.74$. 
The data available to the public corresponds to a downsampled set 
of $10^5$ particles, which corresponds to a particle mass of
$2.24\times 10^{7}$\hMsun.

\begin{figure*}
\begin{center}
  \includegraphics[width=0.49\textwidth]{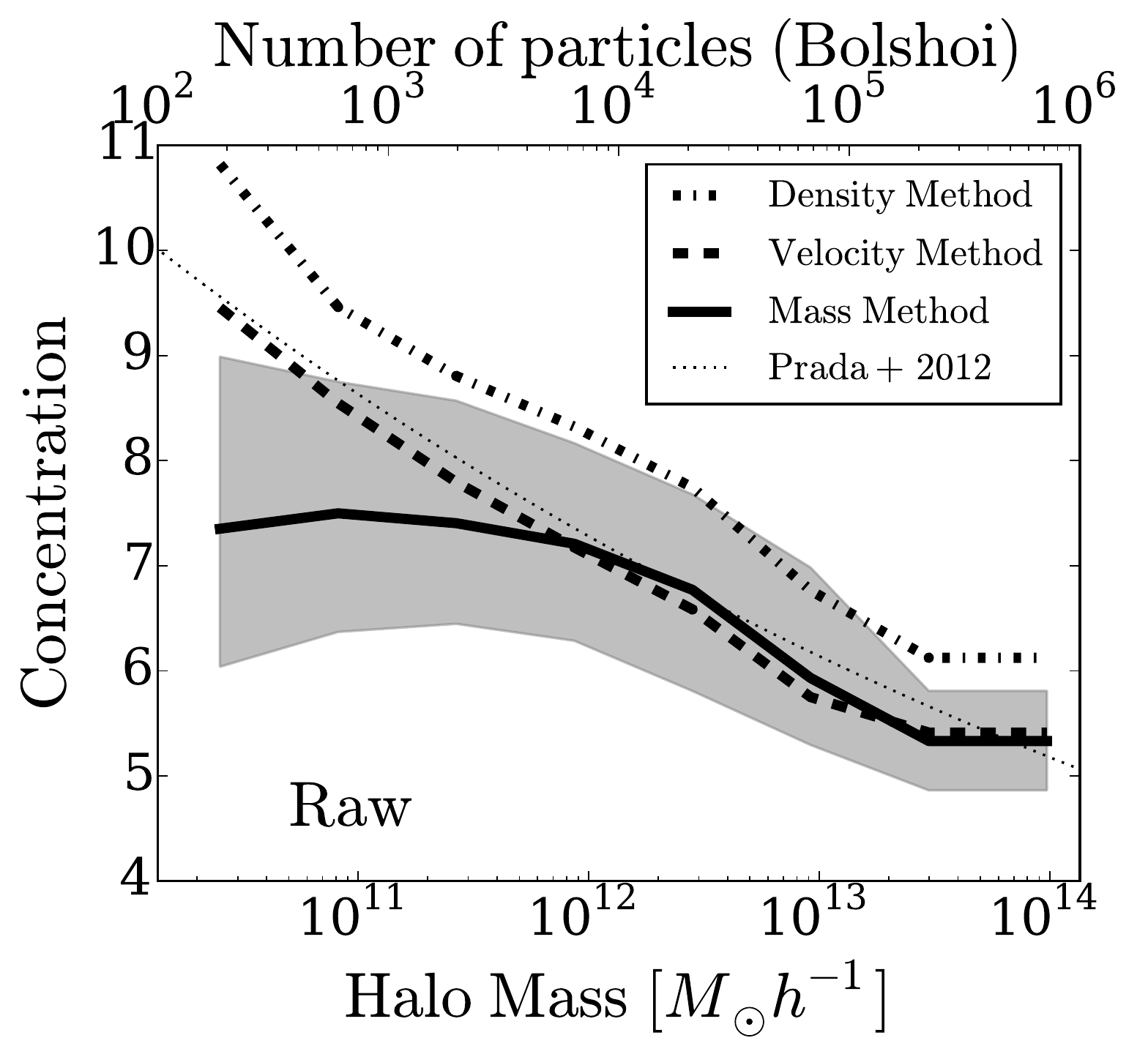}
  \includegraphics[width=0.49\textwidth]{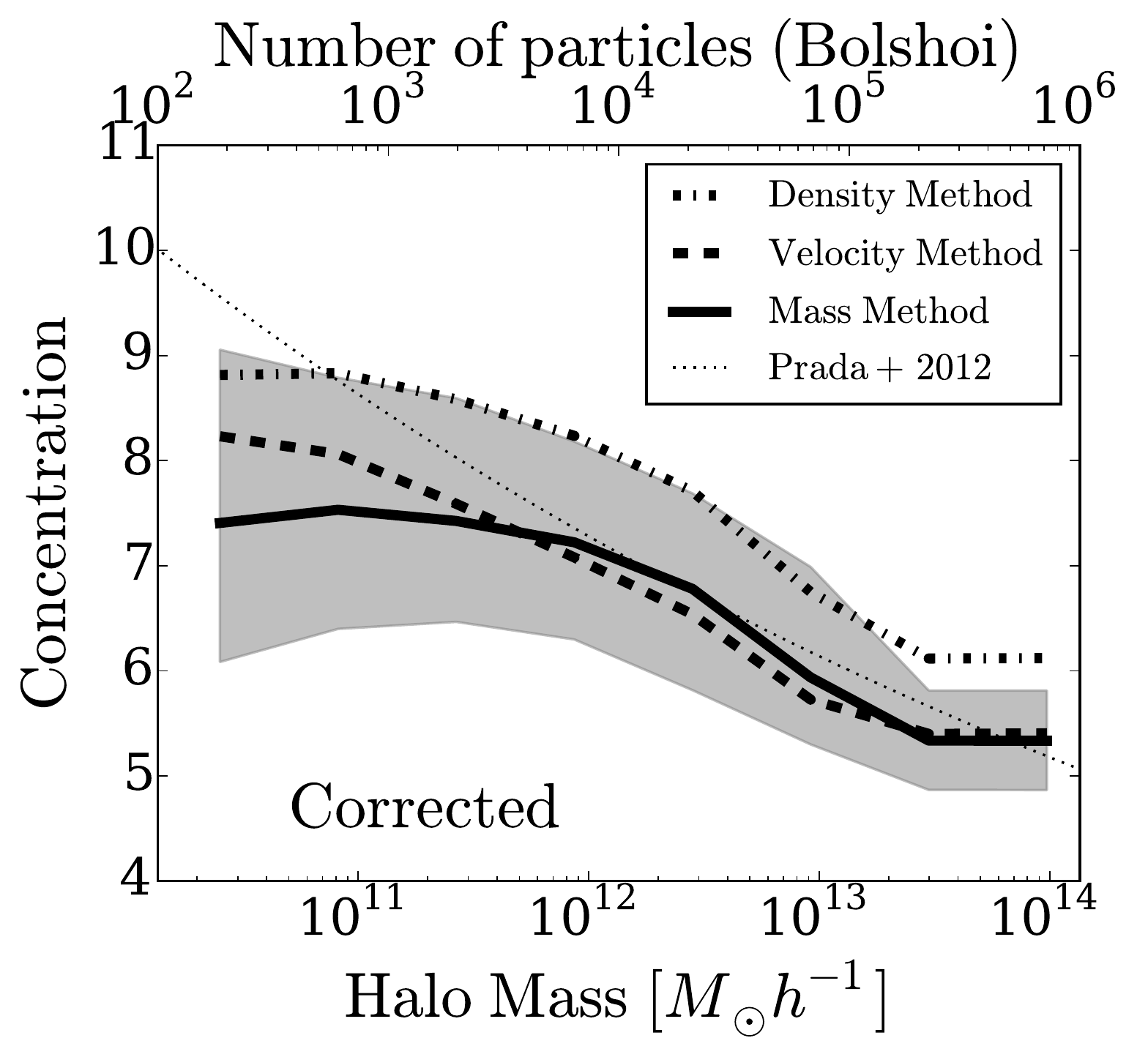}
\end{center}
\vspace{-0.5cm}
\caption{Mass-concentration relationship for the three different
  methods on the Bolshoi data using only relaxed halos.
  The lines correspond to the median concentration values in each mass
  bin.  
  The shaded region presents 10 and 90 per cent spread. 
  The three methods have a similar spread but for clarity we only show
  the spread for the new method.   
  The dotted line corresponds to fits reported by \citep{Prada2012}.
  The left panel shows the raw results coming from each algorithm. 
  The right panel introduces a correction following the results on the
  bias as a function of particle number.
  \label{fig:concentration}} 
\end{figure*}

\section{Results}
\label{sec:results}

\subsection{Bootstrapping to estimate biases}
\label{sec:bootstrapping}

We take halos with at least $10^{5}$ particles and subsample them by
factors of $2$ up to $10^{3}$. 
We measure the concentration at every resampling.
We use a two-sample Kolmogorov-Smirnov (KS) test to  compare the list
of radial distances from each subsample against that of its parent halo.
We find that the resulting p-value distribution is flat.
This confirms that the radial particle distribution in the bootraped
halo is consistent with coming from the distribution given by the
parent halo.  
Why not using different simulations with the same initial conditions
and lower resolutions \citep[i.e.][]{2008MNRAS.391.1685S}?. 
Because we want to be sure that we are only measuring the bias of a
given method as a function of particle number for statistically
identical halos, and not a possible simulation artifact that changes
the halo structure.

For every subsample we keep fixed the virial radius and the center
found for the high resolution halo.
Leaving the virial radius and center free in each bootstrapping
iteration has an effect smaller than $1\%$ in the concentration. 
In the Bolshoi simulation we select 14 massive halos and create 700
subsamples for each one.  
For the Via Lactea simulation the same halo is subsampled 10000
times. 

The average concentration value for the largest number of particles,
$c_{N_{max}}$, provides a baseline to compare all the other results.  
We use the following statistic
\begin{equation}
f_{\rm off} = c_{N} / c_{N_{max}} - 1,
\label{eq:f_off}
\end{equation}
to account for the offset between the concentration at a given
downsampled particle number $c_{N}$ and the baseline $c_{N_{max}}$.

Figure \ref{fig:downsampling} summarizes our results.  
The plot on the left shows the average value of $f_{\rm off}$ as a
function of particle number.  
This can be interpreted as the statistical bias on the concentration
estimate. 
For large enough particle numbers, $N_{s}>4\times 10^3$ the results of
the three algorithms show a bias below the $1\%$ level.
For a lower number of particles the results start to deviate.
At $200$ particles the velocity method overestimates the concentration
by a factor $14\%$ while the density method overestimates it by $20\%$.
Around the same sampling scale, the new algorithm shows a more stable
behaviour underestimating the concentration only by a factor of
$1\%$-$3\%$.

The thin lines on the same panel show a fit to the function 
\begin{equation}
f_{\rm off} = \frac{A}{(1+\log_{10}N_s)^{B}},
\label{eq:f_off_fit}
\end{equation} 
with $A=2842\pm 1900$, $B=7.96\pm 0.54$; $A=239\pm 131$, $B=6.23\pm
0.43$ and $A=-0.46\pm 3.49$, $B=0.79\pm 1.31$ for the density,
velocity and mass method, respectively.

The right panel in Figure \ref{fig:downsampling} shows different
uncertainty results.  The lines show the difference between the 14 and
86 percentiles in the $f_{\rm off}$ distribution at fixed mass.  Each
line corresponds to the three different methods to estimate the
concentration applied to both simulations.  This shows that the
bootstrapping technique can help us to assign a $1\sigma$ uncertainty
to the concentration values at a fixed $N_s$as
\begin{equation}
\sigma_{c} = \frac{0.40}{\sqrt{N_s/200}}.
\end{equation}

The circles in the same Figure show the $1\sigma$ uncertainty on all
the relaxed halos in the Bolshoi simulation sample using the MCMC
results.  
To allow for a fair comparison with the bootstrapping results, this
uncertainty is normalized to the concentration value.
The uncertainty from the bootstrapping experiment provides an upper
bound uncertainty on the concentration estimate for individual halos.

\subsection{Impact on the Mass-Concentration Relationship}

We now inspect the mass-concentration relationship 
results with the three different algorithms.
This can help us to identify possible consequences of the 
biases detected through the bootstrapping experiments.

Figure \ref{fig:concentration} shows the mass-concentration
relationship for the density, velocity and integrated mass method.
The left panel shows the results as they are produced by each of the
algorithms. The thin dashed line marks the trend reported by
\citep{Prada2012} using the velocity method, showing that our
velocity method implementation can reproduce their results. 

The results from the new algorithm follow very
closely the velocity algorithm at high masses ($M_h>10^{12}\hMsun$ or
equivalently for $>4\times10^3$ particles). 
For lower masses there is a difference between the median of the two
methods, but they are still consistent within the statistical
uncertainties.

We hypothesize that the increase in the results for the velocity
and density methods below $4\times 10^{3}$ particles comes from the
systematic bias described in the previous section.  
To test the general consistency of this hypothesis, we correct the
concentration values in the velocity and integrated mass methods by a
factor of $1/(1+f_{\rm  off})$, using the definition in Equation
(\ref{eq:f_off}) and the  
parameters obtained from the data presented in Figure
(\ref{fig:downsampling}).  
The correction brings into good agreement the results between the
velocity/density methods and the new algorithm.

We also notice that the results from the density method have a
systematic $15\%$ offset from the velocity methods.  
This offset was already presented by \cite{Prada2012} for low
concentrations ($c<6$) and high ($M_h>10^{12}\hMsun$) halo masses.  
Recently \citep{Klypin2016} summarized results for the
mass-concentration relationship coming from different methods and
datasets to show that similar systematic offsets are present.
\citep{2014MNRAS.441.3359D} studied the mass-concentration
relationship using the maximum velocity and density methods and did
not report any significant difference. 
However, they implemented a modified version of the velocity algorithm
that bins the particle data, which might explain why they the offset
was not reported.

How do these results impact the most recent mass-concentration estimates? 
\cite{Ludlow2016} and \cite{Klypin2016} estimated the mass-concentration
relation over different suites of cosmological N-body simulations using the
density and velocity methods, respectively. 
Both used halos with at least $5\times 10^{3}$ particles.
This imposes a lower halo mass limit of $\sim 10^{12}$\hMsun (Figure 8
in \cite{Ludlow2016}, Figure 17 in \cite{Klypin2016}) to have robust
estimates. 
This means that their results for individual halos should not be
affected by the bias we report here.
This also leaves open the question about what other methods
can robustly say about the flattening we report below $10^{12}$\Msun
using the new method. 
However, there are other results at lower masses and higher redshifts
\citep[i.e.][]{Prada2012} that should be reconfirmed using higher resolution
simulations as they use halos with only $500$ particles.

\section{Conclusions}
\label{sec:conclusions}

In this letter we used bootstrapping to quantify the biases on
concentration estimates. 
We found that methods commonly used in the literature can overestimate
the concentrations by factors of $15\%$-$20\%$ for halos with $200$
particles, or $7\%$-$10\%$ for halos with $500$ particles.  
This procedure provides a robust technique to quantify the bias in
concentration estimates with the advantage that it works without
having to run new simulations. 

These results motivated us to introduce a new method based on the
integrated mass profile that show a robust performance at low particle
numbers. 
The new algorithm showed a bias  of $< 3\%$ for halos with $200$
particles and less than $1\%$ for halos with $500$ particles or more.  
To keep the bias of the velocity and density methods below $2\%$ only halos
with at least $\sim 4000$ particles should be considered.

The three methods are in broad agreement, within the statistical
uncertainties, concerning their estimates of the  mass-concentration
relationship. 
Some noticeable differences include a $15\%$ systematically higher 
concentrations in the density method compared to the velocity method.
This systematic offset has been reported before with the same dataset 
\citep{Prada2012} and with different simulations \citep{Klypin2016}
without any conclusive explanation for its origin. 
Another difference is that the velocity and integrated mass methods
start to differ for masses below $10^{12}\hMsun$ ($\sim 4000$
particles).   
We found that correcting the mean concentration by the mean bias
factor found through bootstrapping brings these two techniques into
agreement. 

These results show that using the integrated mass profile to estimate the
DM halo concentrations is a tool deserving deeper scrutiny.   
Further tests with larger simulated volumes, varying numerical
resolution, higher redshifts, stacked data and different density
profiles are the next natural step to explore the full potential of
this new method.   

\vspace{0.1cm}

 We acknowledge financial support from Uniandes and Estrategia de
 Sostenibilidad 2014-2015 Universidad de Antioquia.  We thank Tom\'as
 Verdugo, Stefan Gottloeber and Nelson Padilla for their feedback.  We
 thank the anonymous referees for comments that improved the
 presentation of these results.

\bibliographystyle{apj}

\end{document}